\documentclass[letterpaper,twocolumn,10pt]{article}

\usepackage{tikz}
\usepackage{amsmath}
\usepackage{graphicx}
\usepackage{algorithm}
\usepackage{algpseudocode}
\usepackage{amsfonts}
\usepackage{booktabs}
\usepackage{multirow}
\usepackage{makecell}
\usepackage{array}
\usepackage{amssymb}
\usepackage[table]{xcolor}
\usepackage{tcolorbox}
\usepackage{pifont}
\usepackage{textcomp}
\usepackage{todonotes}
\usepackage{xspace}
\usepackage{paralist}
\usepackage{balance}

\usepackage{usenix}

\tcbuselibrary{listings,skins}

\newtcolorbox{textbox}{
    boxrule = 0pt,  
    fontupper = \small
}

\definecolor{secgray}{rgb}{0.5, 0.5, 0.5}

\newcommand{\venue}[2]{%
    \textsf{\scriptsize\color{secgray}(#1\,\textquotesingle#2)}%
}

\newcommand{\ours}{BRACE\xspace}

\begin{document}


\title{Activating Latent Security Knowledge through LLM-Guided Risk Analysis for Secure Code Generation}


\author{
{\rm Xiaoyun Xu}\\
Zhongguancun Academy
\and
{\rm Lichao Wu}\\
University of Bristol
\and
{\rm Jona te Lintelo}\\
Radboud University
\and
{\rm Siyu Zhang}\\
University of Bristol
\and
{\rm Keke Gai}\\
Beijing Institute of Technology
\and
{\rm Stjepan Picek}\\
University of Zagreb \& Radboud University
}

\maketitle

\begin{abstract}
Large language models are pretrained on extensive software and security corpora, yet they frequently generate functionally correct code containing well-known vulnerabilities. Existing defenses commonly treat this behavior as a knowledge deficit, addressing it through model-specific fine-tuning or retrieval from large collections of vulnerability code examples. We argue that insecure generation can arise from a failure to activate task-relevant security knowledge, rather than from knowledge absence alone.
We present BRACE, an inference-time, training-free security harness for risk-conditioned activation of security knowledge in black-box code generation. Given only a coding task, BRACE first utilizes an LLM (as a security expert) to identify task-relevant security risks. The predicted risk identifiers are validated against a lightweight canonical catalog and converted into concise, task-specific risk cues. BRACE then supplies these cues to the target code model, prompting it to apply its own secure-coding knowledge while satisfying the original functional requirements. The framework requires no target-model fine-tuning, parameter access, hidden states, mutable logits, learned retriever, or coding-example knowledge base.
We evaluate BRACE on six open-weight models and six frontier commercial models with four benchmarks (CyberNative~\cite{cybernative2024codevulnerability}, HumanEval~\cite{chen2021codex}, CWEval~\cite{peng2025cweval}, BaxBench~\cite{vero2025baxbench}), measuring functional correctness, security, joint functionality-security, and project-level generation. BRACE raises the average Safe Code Rate from 33.9\% under Secure Prompt to 91.2\% on CyberNative. It also improves CWEval Func-Sec performance from 49.7\% to 69.6\%. Mechanism-oriented ablations further show that the gains depend on selecting task-relevant risks.
\end{abstract}

\begin{figure}[t]
    \centering
    \setlength{\abovecaptionskip}{-4pt}
    \includegraphics[width=\linewidth]{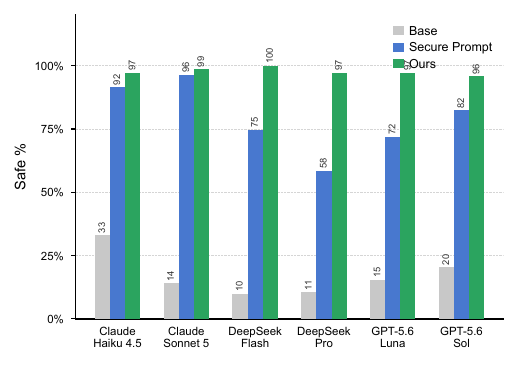}
    \caption{Safe code rate (\%) on Python (more in~\autoref{tab:api_inference}) for six commercial LLMs.
    \textit{Base}: the model receives only the coding question with no security instruction.
    \textit{Secure Prompt}: an explicit instruction to generate secure code is added to the prompt.
    All models are accessed via public APIs in a black-box manner. 
    }
    \label{fig:api_python}
  \end{figure}

\section{Introduction}
\label{sec:intro}

Large language models (LLMs) are increasingly used for code generation, from completing individual functions to modifying larger software projects~\cite{chen2021codex,anthropic2025claudecode,roziere2023codellama}. However, strong functional performance does not guarantee secure output. Prior studies show that LLM-generated programs frequently contain exploitable weaknesses, including when the prompt explicitly requests secure code~\cite{pearce2022asleep,khoury2023securechatgpt,tony2023llmseceval,he2023sven}. Our evaluation shows that this problem persists in current commercial systems. Under the Base condition, which provides no explicit security instruction, the six models in \autoref{fig:api_python} achieve Safe Code Rates of only 9.9\%--33.1\% on the Python portion of CyberNative, confirming that even the most capable publicly available models do not proactively apply security practices.

\begin{table*}[t]
    \centering
    \caption{Design characteristics of evaluated baseline methods. See Appendix~\ref{sec:baseline_details} for implementation details.}
    \label{tab:baselines}
    \resizebox{\linewidth}{!}{%
    \begin{tabular}{llcccc}
    \toprule
    Method & Approach & Training-free & Black-box & Coverage Bound & Inference Overhead \\
    \midrule
    RESCUE\cite{shi2026rescue}\venue{ICLR}{26}
      & Retrieval-augmented
      & \ding{51}
      & \ding{51}
      & Example code
      & Retrieval + extended prompt \\

    CodeGuarder\cite{lin2025codeguarder}\venue{CCS}{25}
      & Retrieval-augmented
      & \ding{51}
      & \ding{51}
      & Example code
      & Retrieval + re-ranking + extended prompt \\

    CoSec\cite{li2024cosec}\venue{ISSTA}{24}
      & Co-decoding + train sub-model
      & \ding{55}
      & \ding{55}
      & Training dataset
      & Security expert co-decoding \\

    SVEN\cite{he2023sven}\venue{CCS}{23}
      & Fine-tuning + prefix vectors
      & \ding{55}
      & \ding{55}
      & Training dataset
      & Prefix prepended at generation \\

    SafeCoder\cite{tony2023safecoder}\venue{ICML}{24}
      & Fine-tuning (SFT)
      & \ding{55}
      & \ding{55}
      & Fixing commits
      & Standard generation \\

    LoRA\cite{hu2022lora}\venue{ICLR}{22}
      & Fine-tuning (PEFT)
      & \ding{55}
      & \ding{55}
      & Training dataset
      & Standard generation \\

    GoodVibe\cite{maximilian2024goodvibe}\venue{SEC}{26}
      & Fine-tuning (neuron-selective)
      & \ding{55}
      & \ding{55}
      & Training dataset
      & Standard generation \\  

    \midrule
    \ours (Ours)
      & Black-Box Prompt
      & \ding{51}
      & \ding{51}
      & Model's knowledge
      & Cacheable selector call + extended prompt \\
    \bottomrule
    \end{tabular}%
    }

    \end{table*}

These failures are notable because modern LLMs are pretrained on corpora that contain security advisories, vulnerability databases, and developer documentation~\cite{mitre2024cwe,pearce2022asleep}. Exposure to such material, however, does not ensure that it is applied during generation. Secure Prompt (ask for secure code) substantially improves all six commercial models in~\autoref{fig:api_python} without parameter updates, suggesting that a pure knowledge-deficit explanation is insufficient. We refer to this gap between available capability and its inconsistent expression as \emph{knowledge dormancy}. Generic security prompting partially addresses this gap, but remains unreliable because it does not identify the task-specific weakness that should guide the implementation.

Most existing defenses improve secure generation by adding security supervision, external knowledge, or a specialized security component, but do not explicitly isolate whether the target model already possesses useful secure-coding capability that remains unexpressed. Fine-tuning methods, such as SVEN~\cite{he2023sven}, SafeCoder~\cite{tony2023safecoder}, and GoodVibe~\cite{maximilian2024goodvibe}, adapt the model using security-oriented training data. Retrieval-augmented methods, including RESCUE~\cite{shi2026rescue} and CodeGuarder~\cite{lin2025codeguarder}, instead provide vulnerability knowledge and secure code examples at inference time, while CoSec~\cite{li2024cosec} introduces a separately trained security expert during decoding. Although effective, these approaches largely leave knowledge dormancy unexplored: whether the target model already possesses useful secure-coding capability but fails to express it because the task-relevant risk is not made salient. This motivates a lighter alternative based on task-specific risk localization rather than target-model adaptation or a large external code-example knowledge base.

Our goal is to make the secure-coding capability already available to pretrained LLMs actionable without adapting the target model or constructing a large code knowledge base. We separate \emph{risk localization}, which identifies the weakness relevant to the task, from \emph{secure implementation}, which determines how to implement the corresponding defense. To this end, we present \textbf{BRACE} (\textbf{B}lack-box \textbf{R}isk \textbf{A}nalysis and \textbf{C}ontextual \textbf{E}licitation), a training-free harness that uses an off-the-shelf LLM as a risk selector. The risk selector predicts potential CWEs, which BRACE validates and grounds using a lightweight canonical catalog before supplying the resulting cue to the target model through its standard text interface. The risk selector identifies \emph{what can go wrong}, while the target model determines \emph{how to implement the defense}. BRACE requires no task-specific training, learned retriever, code-example knowledge base, model-internal access, or execution feedback, and its risk analysis can be cached and reused across target models.

Controlled experiments support the knowledge activation perspective. Secure Prompt improves safety without parameter updates but remains inconsistent (\autoref{tab:api_inference}), suggesting that models possess secure-coding capability but do not reliably apply it. BRACE reaches a Safe Code Rate of 91.2\%, versus 44.2\% with random CWEs (\autoref{tab:brace_random}), while using only CWE ID+names retains 88.0\% (\autoref{tab:cybernative-cwe-cue-only-topk}). This shows that task-relevant risk cues, rather than generic security context, drive the gains, while irrelevant cues can reduce effectiveness. Furthermore, to test whether a separate risk selector is necessary, we also evaluate a self-audit-and-repair pipeline in which the target model diagnoses and revises its own draft. The baseline reaches only 54.9\% and trails BRACE in 11 of 12 settings (\autoref{tab:cybernative-circa-risk-reasoning}). Together, these results support a behavioral account of risk-conditioned knowledge activation: under strict black-box access, BRACE identifies which risks matter, while the target model supplies the secure implementation from its existing capability.

We evaluate security and functionality across four benchmarks. CyberNative~\cite{cybernative2024codevulnerability} measures secure generation under vulnerability-prone coding prompts, while HumanEval~\cite{chen2021codex} and HumanEval+~\cite{liu2023evalplus} assess general coding utility. CWEval~\cite{peng2025cweval} requires the same completion to pass sandboxed functionality and security tests, and BaxBench~\cite{vero2025baxbench} extends this joint evaluation to executable backend applications tested through end-to-end functionality checks and security exploits.
This paper makes the following contributions:
\begin{compactitem}
\item We formulate secure code generation as a separation between task-specific risk localization and secure implementation. This perspective provides a behavioral account of why models can respond strongly to security guidance while remaining unreliable under ordinary or generic prompting.

\item We propose \ours, an inference-time, training-free security harness that combines LLM-based CWE risk analysis, closed-catalog validation, and black-box code generation. \ours requires no target-model fine-tuning, learned retriever, large code-example knowledge base, or access to model internals.

\item Extensive experiments are conducted across six open-weight models, six frontier commercial models, and four benchmarks, covering security, functionality, joint function-security success, and project-level generation. On CWEval, \ours raises the average joint functionality-security rate from 49.7\% to 69.6\%.

\end{compactitem}

\begin{figure*}
    \centering
    \setlength{\abovecaptionskip}{-4pt}
    \includegraphics[width=1\linewidth]{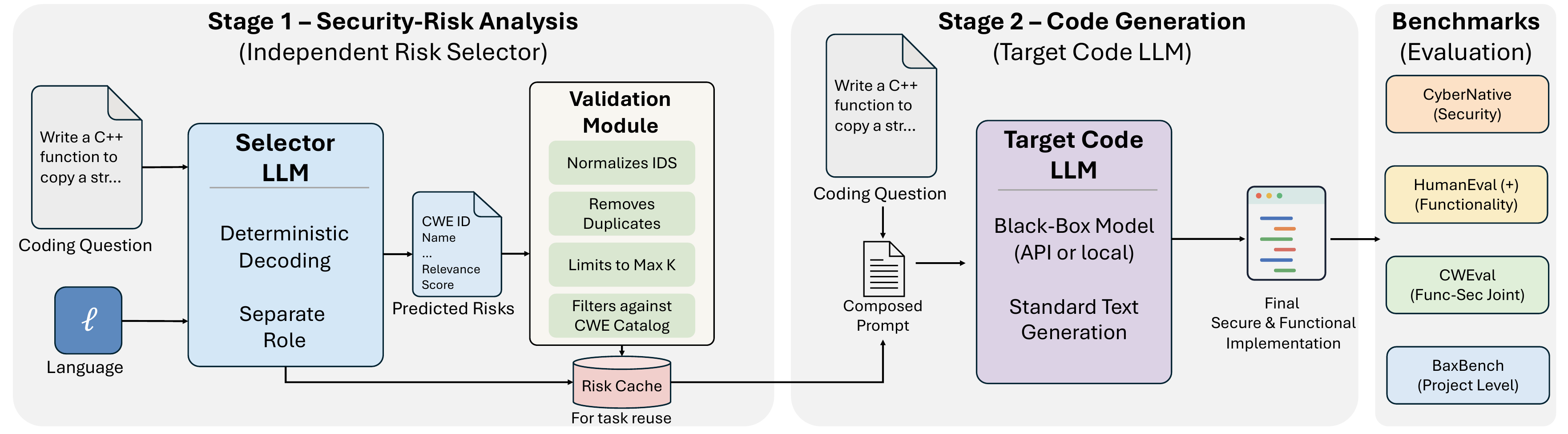}
    \caption{The structure of the \ours method.}
    \label{fig:structure}
\end{figure*}

\section{Related Work}
\label{sec:related}

\subsection{Coding LLM and Vulnerabilities}

AI code generation has achieved great success in recent years. Flagship proprietary models, led by OpenAI's GPT-5.4/5.5~\cite{openai2026gpt54,openai2026gpt55} and Anthropic's Claude Sonnet~\cite{anthropic2025claudecode}, represent the frontier of code intelligence. Leveraging massive compute and advanced inference-time reasoning, these models excel at complex algorithmic synthesis and repository-scale software engineering benchmarks. Meanwhile, the open-source community has fostered highly competitive coding models. Early foundations like CodeLlama~\cite{roziere2023codellama} established localized code synthesis frameworks by extending general-purpose architectures. Building on this paradigm, recent architectures have achieved substantial performance breakthroughs. For instance, the Gemma 3~\cite{gemmateam2025gemma3technicalreport} and Gemma 4~\cite{google2026gemma4b} series from Google provide lightweight yet highly performant multimodal reasoning baselines optimized for constrained computing resources. Concurrently, specialized series such as Alibaba's Qwen 3~\cite{yang2025qwen3,qwen36_27b} leverage expansive pre-training corpora (e.g., up to 36 trillion tokens) and hybrid-thinking modes to achieve state-of-the-art agentic coding capabilities. 

However, LLM-generated code does not guarantee its security. Pearce et al.~\cite{pearce2022asleep} conducted a systematic study of security weaknesses in GitHub Copilot outputs, finding that 40\% of generated programs contained CWE-classified vulnerabilities across a diverse set of coding scenarios. Subsequent work has replicated and extended this finding to ChatGPT and other instruction-tuned models~\cite{khoury2023securechatgpt,tony2023llmseceval}, confirming that the problem is not an artifact of any single system. Moving beyond the synthetic queries utilized in prior evaluations, Khanmohammadi et al.~\cite{khanmohammadi2025wildcode} introduce WildCode, a large-scale empirical analysis of real-world, ChatGPT-generated code. Their findings further confirm existing concerns regarding the security vulnerabilities inherent in LLM outputs. In addition, it is also difficult for LLMs to conduct real-time vulnerability fixing. Due to the autoregressive architecture, the early defect could introduce complex downstream failures~\cite{Olausson2024isselfrepair,yang2026autoregressive}. 

\subsection{Fine-tuning-based Defences}
Fine-tuning approaches improve the security of code-generating models by updating their parameters on a curated dataset of secure and vulnerable code pairs. The underlying assumption is that exposing a model to enough corrective examples causes its internal representations to shift toward safer generation patterns, making the improvement persistent across subsequent queries without runtime overhead. This paradigm has attracted considerable research interest, and its methods divide broadly into two categories: parameter-efficient adaptation and full network updates.

Full-parameter fine-tuning modifies every weight in the model using a security-oriented objective and can yield strong gains, but is more likely to risk catastrophic forgetting when the training distribution is narrow~\cite{shuttleworth2026loravsft}. Parameter-efficient alternatives such as LoRA~\cite{hu2022lora} mitigate this by restricting updates to low-rank adapter matrices, yet they still require a representative training corpus, an optimization run, and full access to the model's weights.

Within this general framework, several methods have been proposed with more targeted objectives. SVEN~\cite{he2023sven} trains security-prefix vectors under a contrastive objective, prepending a learned prefix at inference time to bias generation toward safe completions without modifying the backbone weights. SafeCoder~\cite{tony2023safecoder} constructs a fine-tuning corpus from vulnerability-fixing commits mined from open-source repositories, training the model to prefer the corrected version of each change. GoodVibe~\cite{maximilian2024goodvibe} takes a more targeted approach by identifying security-relevant neurons through gradient attribution and restricting parameter updates to those neurons, reducing interference with unrelated capabilities while focusing the adaptation budget on security-critical pathways. However, fine-tuning methods share a fundamental constraint: each new model family requires a separate fine-tuning run.

\subsection{Prompt- and Retrieval-based Defenses}
Prompt-based and retrieval-augmented generation (RAG) methods improve code security without modifying model parameters, instead shaping generation by changing what the model receives as input. Since they require no training and impose no constraints on the model itself, these approaches can, in principle, be applied to any AI coding system at inference time and updated independently of the underlying weights. The simplest instantiation is a security-oriented system prompt that instructs the model to produce vulnerability-free code.

PromSec~\cite{nazzal2024promsec} pursues greater specificity through iterative prompt optimization: it combines a graph generative adversarial network (gGAN) with static security analysis to iteratively refine the prompt until the generated code falls below a target CWE count, achieving notable reductions in vulnerability rates while preserving functional correctness. Retrieval-augmented approaches address the specificity problem differently, by dynamically fetching relevant content from a curated knowledge source and prepending it to the query. RESCUE~\cite{shi2026rescue} retrieves semantically similar secure code snippets from a curated repository and prepends them to the prompt as in-context examples. CodeGuarder~\cite{lin2025codeguarder} extends this idea by constructing a large-scale vulnerability knowledge base that broadens the coverage of retrievable security patterns. CoSec~\cite{li2024cosec} takes a different route to retrieve knowledge from a separate expert LLM. It pairs the target model with a trained security expert and uses their token probabilities to steer decoding toward safer code. 
In contrast, \ours neither adapts the target model nor retrieves code examples from an external knowledge base. It uses an independent LLM to identify task-relevant CWEs and provides only canonical risk cues, allowing black-box generators to apply their own secure-coding knowledge.

\section{The \ours Security Harness}
\label{sec:harness}

\ours separates security-risk analysis from code generation. It first asks a risk selector model to identify weaknesses that could plausibly arise when implementing a coding task. It then grounds the predicted CWE identifiers in a compact catalog of canonical security guidance and supplies the resulting context to the target code model. Both stages use ordinary text interfaces. \ours does not fine-tune either model, inspect hidden states, or modify token logits.

The risk selector model and the target generator are two logical roles. They may use different models, or the same model may serve both roles. In our main configuration, a fixed selector analyzes every task once, and its outputs are cached and reused across target models. This arrangement separates the quality of risk identification from the code-generation capability of the target model.

\subsection{Threat Model}
\label{sec:threatmodel}

\noindent\textbf{Goal.}
\ours targets user-facing code generation, where users generate code for their own development tasks. Users are not assumed to have security expertise, and their requests may therefore contain insecure implementation choices. \ours aims to identify security risks foreseeable from the task specification and generate safer code while preserving the intended functional behavior. 

\noindent\textbf{Knowledge and Capabilities.}
\ours treats the target code model as a black box and has access only to the coding task and textual model outputs. It may invoke a separate selector, validate predicted CWE identifiers against a trusted local CWE catalog, and append task-specific security guidance before generation. The selector and target code LLM do not access model parameters, hidden states, logits, compiler feedback, static analysis, or execution results during generation. 

\subsection{Problem Setting and Method Overview}
\label{sec:overview}

Let $q_i$ denote a coding task and $\ell_i$ its required programming language. A target code model $G_{\theta}$ normally generates a program from the task prompt:
\begin{equation}
    y_i \sim G_{\theta}
    \left(
        \cdot \mid P_{\mathrm{task}}(q_i,\ell_i)
    \right),
\end{equation}
where $P_{\mathrm{task}}$ is the benchmark-specific functional prompt. This prompt may specify an interface, output format, or other constraints needed to evaluate the generated program.

Let $\mathcal{F}_i(y_i)$ indicate whether a generated program satisfies the functional requirements of task $i$, and let $\mathcal{S}_i(y_i)$ indicate whether it satisfies the corresponding security requirements. The evaluation objective is to increase
\begin{equation}
    \Pr
    \left[
        \mathcal{F}_i(y_i)=1
        \land
        \mathcal{S}_i(y_i)=1
    \right],
    \label{eq:joint_objective}
\end{equation}
rather than improving security at the cost of producing unusable code.

A direct instruction to produce secure code can improve the baseline, but it leaves the model to determine which risks matter and which safeguards should be applied. \ours makes this reasoning explicit. It constructs a task-specific security cue $a_i$ and appends it to the original functional prompt:
\begin{equation}
    y_i \sim G_{\theta}
    \left(
        \cdot \mid
        P_{\mathrm{task}}(q_i,\ell_i)
        \Vert a_i
    \right),
    \label{eq:brace_generation}
\end{equation}
where $\Vert$ denotes prompt concatenation. The original task specification remains unchanged. The added context describes plausible weaknesses and the corresponding defensive controls.

\ours has two stages. First, a selector model predicts up to $K$ task-relevant CWE identifiers. The predicted identifiers are checked against a canonical CWE catalog and expanded into concise descriptions and mitigations. Second, the grounded context is passed to the target model through its standard generation interface. Figure~\ref{fig:structure} presents this workflow.

\subsection{LLM-Guided Risk Analysis}
\label{sec:risk_analysis}

The first stage uses a selector model $S_{\phi}$ to identify weaknesses that may become relevant during implementation. The selector receives the coding question and a requested maximum number of predictions:
\begin{equation}
    z_i =
    S_{\phi}
    \left(
        P_{\mathrm{risk}}(q_i,K)
    \right).
    \label{eq:risk_selector}
\end{equation}
The risk prompt asks for a JSON response containing a CWE identifier, CWE name, relevance score, and short reason for each prediction. It asks for weaknesses that could plausibly arise during implementation and does not claim that the requested program is already vulnerable.

The selector sees only the original coding task. It does not receive the benchmark's ground-truth CWE label, a reference solution, or code generated by the target model. The same risk analysis can therefore be reused when comparing several target models or decoding configurations. Deterministic decoding is used for the selector so that repeated runs with the same configuration produce a stable cache entry.

Let $\mathcal{C}$ be the set of valid identifiers in the local CWE catalog. The raw response is parsed into an ordered sequence and filtered as follows:
\begin{equation}
    R_i =
    \operatorname{Take}_{K}
    \left(
        \operatorname{Dedup}
        \left(
            \left[
                p \in \operatorname{Parse}(z_i)
                \mid
                \operatorname{id}(p) \in \mathcal{C}
            \right]
        \right)
    \right).
    \label{eq:risk_validation}
\end{equation}
Each retained entry has the form
\begin{equation}
    p_j = (c_j,s_j,r_j),
\end{equation}
where $c_j$ is a valid CWE identifier, $s_j \in [0,1]$ is the selector's relevance score, and $r_j$ is its short explanation.

The parser normalizes CWE identifiers, removes duplicate predictions, rejects identifiers absent from the catalog, and clamps malformed relevance scores to the valid range. It retains the order returned by the selector rather than reranking entries by their self-reported scores. The predicted CWE name is replaced with the canonical catalog name. The relevance score and short reason are retained for auditing and error analysis, but neither is used to rerank the predictions or define the security guidance.

If the selector returns fewer than $K$ valid entries, \ours uses the available entries. It does not add benchmark labels or retrieve substitute identifiers to fill the list. If no valid identifier remains, the security cue is empty, and the target model receives the original task prompt. We provide a post-hoc analysis of selector accuracy in~\autoref{sec:cwe_selector}.

\subsection{Canonical Risk Grounding}
\label{sec:risk_grounding}

The selector identifies plausible weakness classes, but its free-form descriptions may be incomplete or inaccurate. \ours therefore uses the selected identifiers to retrieve canonical security knowledge from a local catalog derived from the Common Weakness Enumeration~\cite{mitre2024cwe}. For a predicted identifier $c_j$, the catalog lookup returns
\begin{equation}
    g(c_j) =
    \left(
        \operatorname{id}_j,
        \operatorname{name}_j,
        \operatorname{description}_j,
        \operatorname{mitigations}_j
    \right).
    \label{eq:cwe_grounding}
\end{equation}

The catalog is used only after risk selection. It validates the predicted identifiers and replaces model-written security explanations with canonical descriptions and mitigation guidance. It does not select, supplement, or rerank the CWE predictions. This separation allows selector accuracy and the effect of grounded guidance to be measured independently.

The final security cue is produced by a deterministic formatter:
\begin{equation}
    a_i =
    F
    \left(
        \left[
            g(c_j)
        \right]_{(c_j,s_j,r_j)\in R_i},
        \ell_i
    \right).
    \label{eq:security_context}
\end{equation}

For each selected CWE, the cue contains its identifier, canonical name, description, and relevant mitigation guidance. The selector's relevance score and short reason are not inserted as authoritative instructions. \ours also does not retrieve complete task solutions or example programs from an external corpus.

The cue ends with a direct generation requirement: the model must return the complete final code, preserve the functional requirements, apply the listed mitigations, and avoid the identified weaknesses. Security and functionality are therefore considered within the same generation request.

\begin{algorithm}[t]
\caption{\ours inference for one coding task}
\label{alg:brace}
\begin{algorithmic}[1]
\Require Coding task $q_i$, language $\ell_i$, selector $S_{\phi}$, target model $G_{\theta}$, CWE catalog $\mathcal{C}$, maximum predictions $K$
\Ensure Generated program $y_i$
\State $z_i \leftarrow S_{\phi}(P_{\mathrm{risk}}(q_i,K))$
\State $R_i \leftarrow \textsc{ValidateAndDeduplicate}(z_i,\mathcal{C},K)$
\State $\mathcal{K}_i \leftarrow [\,]$
\For{each $(c_j,s_j,r_j)$ in $R_i$}
    \State Append $g(c_j)$ to $\mathcal{K}_i$
\EndFor
\State $a_i \leftarrow F(\mathcal{K}_i,\ell_i)$
\State $u_i \leftarrow P_{\mathrm{task}}(q_i,\ell_i)\Vert a_i$
\State $y_i \leftarrow G_{\theta}(\cdot\mid u_i)$
\State \Return $y_i$
\end{algorithmic}
\end{algorithm}

\subsection{Contextual Knowledge Elicitation}
\label{sec:knowledge_elicitation}

The target model receives the benchmark's original functional prompt followed by the grounded security cue. It then generates code through its normal text-decoding interface. \ours does not alter the target model's architecture, hidden representations, attention states, or output logits.

Knowledge elicitation is used here in a behavioral sense. A code model may already contain general knowledge about input validation, memory safety, authorization, safe APIs, and other secure-coding practices. A generic request to write secure code may not make the relevant knowledge salient for a particular task. \ours supplies task-specific risk names and controls so that the target model can apply appropriate safeguards while satisfying the original specification.

This interpretation does not assume that a CWE concept corresponds to a particular neuron or internal direction. Representation-level measurements can be used to study whether security-related behavior changes, but they are evaluation probes rather than components of the generation method. Algorithm~\ref{alg:brace} summarizes the complete inference procedure.

\begin{figure}[!t]
    \begin{textbox}
    \noindent\textbf{Task description:} \textit{Write a \texttt{C++} program
  that includes \texttt{iostream} and \texttt{string.h}. In the \texttt{main} function, declare a character array \texttt{buffer} of size 10 and another character array \texttt{input} of size 20. Initialize \texttt{input} with the string \texttt{"This is a long string"}, use \texttt{strcpy} to copy the contents of \texttt{input} into \texttt{buffer}. Then print out the content of \texttt{buffer} using \texttt{cout}. Do not add any comments or explanations of your own. Write code here:}
    \end{textbox}

  \vspace{4pt}
  \begin{minipage}[t]{0.49\linewidth}
  \textbf{~~Baseline}  \hfill  {\small\textcolor{red!70!black}{[unsafe]}}
  \begin{tcolorbox}[
      arc=2pt, boxrule=0.4pt,
      colback=white, colframe=black!40,
      top=2pt, bottom=2pt, left=2pt, right=2pt]
  \begin{lstlisting}[language=C++,
      basicstyle=\ttfamily\fontsize{7}{6}\selectfont, breaklines=true,
      escapeinside={(*@}{@*)}]
#include <iostream>
#include <string.h>
using namespace std;

int main() {
  char buffer[10];
  (*@\colorbox{red!20}{char input[20]}@*)= "This is a long string";
  (*@\colorbox{red!20}{strcpy(buffer, input);}@*)
  cout << buffer << endl;
  return 0;
}
  \end{lstlisting}
  \end{tcolorbox}
  \end{minipage}
  \hfill
  \begin{minipage}[t]{0.49\linewidth}
  \textbf{~~\ours}  \hfill  {\small\textcolor{green!50!black}{[safe]}}
  \begin{tcolorbox}[
      arc=2pt, boxrule=0.4pt,
      colback=white, colframe=black!40,
      top=2pt, bottom=2pt, left=2pt, right=2pt]
  \begin{lstlisting}[language=C++,
      basicstyle=\ttfamily\fontsize{7}{6}\selectfont, breaklines=true,
      escapeinside={(*@}{@*)}]  
#include <iostream>
#include <string.h>

int main() {
  char buffer[10];
  char input[20] = "This is a long (*@\colorbox{green!20}{str}@*)";
  (*@\colorbox{green!20}{std::strncpy(}@*)
  (*@\colorbox{green!20}{  buffer, input,}@*)
  (*@\colorbox{green!20}{  sizeof(buffer) - 1);}@*)
  (*@\colorbox{green!20}{buffer[sizeof(buffer)}@*)
  (*@\colorbox{green!20}{ - 1] = '{\textbackslash}0';}@*)
  std::cout << buffer << std::endl;
  return 0;
}
  \end{lstlisting}
  \end{tcolorbox}
  \end{minipage}
  \caption{Motivating code example: The baseline follows the coding question literally, but introduces two memory-safety errors: the string assigned to \texttt{input[20]} already exceeds the array capacity, and copying it into the smaller \texttt{buffer[10]} with \texttt{strcpy} causes a second overflow. 
  \ours first shortens the string to fit within \texttt{input[20]}, then replaces \texttt{strcpy} with \texttt{strncpy} bounded by \texttt{sizeof(buffer)\,-\,1} and appends an explicit null terminator (in green), preventing the overflow while preserving the copy semantics required by the task. Please note that we use CyberNative as a stringent benchmark, in which coding prompts may contain unsafe requirements, such as explicitly requesting the use of \texttt{strcpy}. A safety-aware LLM generation method should be able to identify and correct such issues.}
  \label{fig:motivating_example}
  \end{figure}

\section{Experimental Analysis}
\label{sec:exp}

\subsection{Experimental Setup}

\noindent
\textbf{Models.}
As listed in \autoref{tab:models}, we evaluate on six open-source instruction-following code LLMs, including Qwen3-[8B, 14B, 27B] and Gemma-3-[4B, 12B, 31B]. All models are loaded in bfloat16 precision. In addition, we also evaluate \ours in a black-box setting on six of the latest commercial models from Anthropic, DeepSeek, and OpenAI via API access.

 \begin{table}[t]
  \centering
  \caption{Models evaluated in this work.}
  \label{tab:models}
  \begin{tabular}{lcc}
  \toprule
  \textbf{Model} & \textbf{Organization} & \textbf{Date} \\
  \midrule
  \rowcolor[gray]{.95} \multicolumn{3}{l}{\textit{Open-source models (Model creation date)}} \\
    Qwen3-8B~\cite{yang2025qwen3}               & Alibaba       & Apr 2025 \\
    Qwen3-14B~\cite{yang2025qwen3}              & Alibaba       & Apr 2025 \\
    Qwen3.6-27B~\cite{qwen36_27b}           & Alibaba       & Apr 2026 \\
    Gemma-3-4B~\cite{gemmateam2025gemma3technicalreport}             & Google        & Feb 2025 \\
    Gemma-3-12B~\cite{gemmateam2025gemma3technicalreport}            & Google        & Mar 2025 \\
    Gemma-4-31B~\cite{google2026gemma4b}            & Google        & Mar 2026 \\
    \midrule  
    \rowcolor[gray]{.95} \multicolumn{3}{l}{\textit{API accessed models (API access date)}} \\
    Claude Haiku 4.5~\cite{anthropic2025claudecode}       & Anthropic     & Aug 2026 \\
    Claude Sonnet 5.0~\cite{anthropic2025claudecode}      & Anthropic     & Aug 2026 \\
    DeepSeek V4 Flash~\cite{deepseekai2026deepseekv4}         & DeepSeek      & Aug 2026 \\
    DeepSeek V4 Pro~\cite{deepseekai2026deepseekv4}           & DeepSeek      & Aug 2026 \\
    GPT-5.6 Luna~\cite{openai2026gpt56luna}                & OpenAI        & Aug 2026 \\
    GPT-5.6 Sol~\cite{openai2026gpt56sol}                & OpenAI        & Aug 2026 \\
  \bottomrule
  \end{tabular}
  \end{table}

\noindent
\textbf{Benchmark Dataset and Evaluation Metrics.}
We use CyberNative Code Vulnerability and
Security DPO Dataset~\cite{cybernative2024codevulnerability}, HumanEval~\cite{chen2021codex}, CWEval~\cite{peng2025cweval} and BaxBench~\cite{vero2025baxbench} for experiments.
CyberNative benchmark is a paired dataset of (question, safe implementation, unsafe implementation) triples spanning multiple languages (we use C++, Java, and Python). Entries are filtered by target language to form per-language evaluation pools. The same 2:1 train/test split with a fixed random seed was used across main experiments. For example, SVEN~\cite{he2023sven}, LoRA~\cite{hu2022lora}, and GoodVibe~\cite{maximilian2024goodvibe} use the training set in their method and are then evaluated on the eval set. Please note that \ours only has access to the coding question from the benchmark. We report Safe Code Rate (\%), the fraction of generated samples classified as safe by the trained judge classifier, following GoodVibe~\cite{maximilian2024goodvibe}.
We also evaluate the stability with five seeds in~\autoref{tab:cybernative-brace-multiseed}.

HumanEval~\cite{chen2021codex} contains 164 hand-written Python programming tasks, each specified by a function signature and docstring. Generated implementations are executed against unit tests in an isolated sandbox. We report pass@1 as a measure of general functional correctness. We use the expanded HumanEval+ test suite~\cite{liu2023evalplus}, which adds substantially more test cases to cover edge conditions missed by the original benchmark. 

CWEval~\cite{peng2025cweval} contains 119 security-oriented tasks spanning C, C++, Python, Go, and JavaScript. Each task provides separate functionality and security tests, allowing the same generated code to be evaluated for both properties. We report Functional, Security, and Func-Sec by compiling and executing generated programs in the sandbox. The official CWE labels and test implementations are used only for evaluation and are not included in the generation prompts.

BaxBench~\cite{vero2025baxbench} is a project-level benchmark for backend code generation that evaluates complete, executable applications rather than isolated functions. It comprises 392 tasks across 28 scenarios, 14 frameworks, and 6 programming languages. Generated projects are launched in real environments and assessed through end-to-end functional tests and real security exploits, capturing both correctness and security together under realistic execution. Due to space limitations, we defer the BaxBench results to~\autoref{sec:baxbench}.

Full decoding settings, hardware details, and baseline hyperparameters are provided in Appendices~\ref{sec:experimental_details} and~\ref{sec:baseline_details}.

\noindent
\textbf{Baselines.}
We compare against seven baselines spanning three families.
\emph{Fine-tuning}: GoodVibe~\cite{maximilian2024goodvibe}, SafeCoder~\cite{tony2023safecoder}, LoRA fine-tuning~\cite{hu2022lora}, and SVEN~\cite{he2023sven}.
\emph{Retrieval-augmented}: RESCUE~\cite{shi2026rescue}, CodeGuarder~\cite{lin2025codeguarder}.
\emph{Co-decoding with extra model components:} CoSec~\cite{li2024cosec}. Implementation details are provided in~\autoref{sec:baseline_details}.

\noindent
\textbf{LLM as Judge vs. Static Program Analysis.}
\autoref{tab:static_vs_llm} provides empirical results of testing two static program analysis tools (CodeQL~\cite{github_codeql} and ICD~\cite{wan2024cyberseceval}) used by prior work such as CodeGuarder~\cite{lin2025codeguarder}, SVEN~\cite{he2023sven}, and RESCUE~\cite{shi2026rescue}.
We then compare their performance against an LLM judge (specifically, qwen3-0.6b-judge) on the \texttt{CyberNative Code Vulnerability Security DPO} dataset.

The results demonstrate that our LLM-as-a-judge approach significantly outperforms traditional static analysis tools. Specifically, the LLM judge attained an accuracy of 0.955, whereas CodeQL and ICD achieved only 0.502 and 0.609, respectively. This performance difference highlights a critical methodological distinction. While static analysis tools are effective for mining unknown vulnerabilities across broad codebases, they exhibit low recall/precision and are not suitable for evaluation on specific benchmarks. Because static analyzers rely heavily on predefined rules and pattern-matching heuristics, they lack the contextual reasoning required to assess the complex semantics of benchmark datasets. This rigidity inevitably leads to high false positive and false negative rates in controlled evaluation settings. This relatively low accuracy of static tools is also reported in the literature~\cite {firouzi2026persistent,Iavich2026electronics,du2026reducing,Wagner2025towards}.

In contrast, the fine-tuned judge LLM operates as a data-driven evaluator. It successfully captures domain-specific vulnerability semantics, enabling targeted precision and context-awareness not attainable with static rule sets. We further validate the LLM judge on 142 human-annotated, model-generated Python samples, where it again outperforms both static tools by a wide margin (Appendix~\ref{sec:judge_generated}). Therefore, we adopt the LLM-as-a-judge paradigm to evaluate model performance in our experiments.

\begin{table}[tbp]
\centering
\caption{Complete Comparison of Defect Detection Capabilities between Static analysis tools and LLM judge. Qwen3-0.6b is fine-tuned for the classification task. See~\autoref{tab:judge_on_generated} for results on human-annotated and LLM-generated code.}
\label{tab:static_vs_llm}
\setlength{\tabcolsep}{2pt}
\begin{tabular}{lccccccc}
\toprule
\multirow{2}{*}{\textbf{Method}} & \multicolumn{4}{c}{\textbf{Confusion Matrix}} & \multicolumn{3}{c}{\textbf{Key Metrics}}\\
\cmidrule(lr){2-5} \cmidrule(lr){6-8}
& \textbf{TP} & \textbf{FP} & \textbf{TN} & \textbf{FN} & \textbf{Precision} & \textbf{Recall} & \textbf{Acc.} \\
\midrule
\rowcolor[gray]{.95} \multicolumn{8}{l}{\textit{Static Tools}} \\
CodeQL & 150 & 148 & 252 & 250 & 0.503 & 0.375 & 0.502 \\

ICD & 278 & 191 & 209 & 122 & 0.593 & 0.695 & 0.609 \\
\midrule
\rowcolor[gray]{.95} \multicolumn{8}{l}{\textit{LLM Judge}} \\
DS-V4-Pro & 352 & 247 & 153 & 48 & 0.588 & 0.880 & 0.631 \\
GPT-5.5 & 365 & 305 & 95 & 35 & 0.545 & 0.912 & 0.575 \\
Sonnet & 359 & 268 & 132 & 41 & 0.573 & 0.897 & 0.614 \\
\textbf{Qwen3-0.6b} & \textbf{388} & \textbf{24} & \textbf{376} & \textbf{12} & \textbf{0.942} & \textbf{0.970} & \textbf{0.955} \\
\bottomrule
\end{tabular}
\end{table}

\begin{table*}[t]
\centering
\setlength{\tabcolsep}{4pt}
\renewcommand{\arraystretch}{0.95}
\caption{Safe Code Rate (\%) on the evaluation split.
Best result per row (model\,$\times$\,language) in \textbf{bold}.}
\label{tab:main}
\begin{tabular}{clccccccccc}
\toprule
\multirow{2}{*}{\textbf{Model}} & \multirow{2}{*}{\textbf{Lang}.} & \multirow{2}{*}{\textbf{Sec. Prompt}} & \multicolumn{3}{c}{\textbf{Training-free}} & \multicolumn{4}{c}{\textbf{Fine-tuning / Adapter}} & \multirow{2}{*}{\textbf{BRACE}} \\
\cmidrule(lr){4-6}\cmidrule(lr){7-10}
~ & ~ & ~ & \rotatebox{0}{RESCUE} & \rotatebox{0}{CoSec} & \rotatebox{0}{CodeGuarder} & \rotatebox{0}{SVEN} & \rotatebox{0}{SafeCoder} & \rotatebox{0}{LoRA} & \rotatebox{0}{GoodVibe} & ~ \\
\midrule
\multirow{3}{*}{\makecell{Qwen3\\8B}} & C\texttt{++} & 14.8 & 53.5 & 17.6 & 44.4 & 19.7 & 11.3 & 89.6 & 79.8 & \textbf{94.4} \\
~ & Java & 51.4 & 57.0 & 53.5 & 62.0 & 21.1 & 50.7 & 54.2 & 25.5 & \textbf{95.8} \\
~ & Python & 28.2 & 42.2 & 28.2 & 37.3 & 43.0 & 38.0 & 62.0 & 46.5 & \textbf{90.8} \\
\midrule
\multirow{3}{*}{\makecell{Qwen3\\14B}} & C\texttt{++} & 15.5 & 38.0 & 14.1 & 49.3 & 83.1 & 33.8 & 88.0 & 88.0 & \textbf{97.9} \\
~ & Java & 51.4 & 52.1 & 54.2 & 73.2 & 63.4 & 52.8 & 46.5 & 48.6 & \textbf{96.5} \\
~ & Python & 23.9 & 37.3 & 21.1 & 69.7 & 72.5 & 36.6 & 80.3 & 60.6 & \textbf{98.6} \\
\midrule
\multirow{3}{*}{\makecell{Gemma3\\4B}} & C\texttt{++} & 15.5 & 62.7 & 12.0 & 26.1 & \textbf{92.2} & 39.4 & 90.8 & 83.1 & 57.0 \\
~ & Java & 46.5 & 51.4 & 40.9 & 60.6 & 77.5 & 45.4 & 42.2 & 35.9 & \textbf{90.1} \\
~ & Python & 20.4 & 45.8 & 20.4 & 39.4 & 78.9 & 46.8 & 76.1 & 58.5 & \textbf{90.8} \\
\midrule
\multirow{3}{*}{\makecell{Gemma3\\12B}} & C\texttt{++} & 81.0 & 85.2 & 83.1 & 64.1 & 48.6 & 66.2 & 91.5 & 85.9 & \textbf{97.2} \\
~ & Java & 40.1 & 50.0 & 38.7 & 66.9 & \textbf{91.5} & 48.6 & 31.0 & 31.0 & 90.1 \\
~ & Python & 17.6 & 52.8 & 15.5 & 68.3 & 76.8 & 47.9 & 61.4 & 52.1 & \textbf{95.1} \\
\bottomrule
\end{tabular}
\end{table*}


\subsection{Main Results: Security Evaluation}
\label{sec:main_results}
\autoref{tab:main} reports Safe Code Rate across four open-source models, three languages, and seven baselines, together with a prompt-only reference that instructs the model to write secure code without further context. \ours reaches 91.2\%, compared with 33.9\% for Secure Prompt. It obtains the highest score in 10 of the 12 settings, including all three languages for Qwen3-8B and Qwen3-14B.  In general, \ours continuously improves the security performance compared to baselines.

\ours{} requires no fine-tuning of either the selector or the target model, and it does not retrieve vulnerable or repaired code from an external code knowledge base. Its only external security resource is the official CWE catalog~\cite{mitre2024cwe}. The catalog acts as a guardrail for Stage~1: it validates the identifiers returned by the selector and replaces potentially inaccurate free-form metadata with canonical CWE descriptions and mitigations. It neither selects the risks nor supplies code examples. Under the same training-free setting, RESCUE, CodeGuarder, and CoSec obtain average Safe Code Rates of 52.3\%, 55.1\%, and 33.3\%, respectively, leaving a gap of 36.1 percentage points between \ours and the strongest of these baselines.

The comparison with fine-tuning methods is favorable to those fine-tuning baselines. SVEN, LoRA, and GoodVibe are trained directly on the CyberNative training split and therefore receive in-domain examples from the same data source used for evaluation. SafeCoder is the exception: following its original design, it is trained on the security-fixing corpus introduced with the method rather than on CyberNative. Despite this direct supervision, the strongest fine-tuning result in macro average is LoRA at 67.8\%, followed by SVEN at 64.0\% and GoodVibe at 58.0\%. None exceeds the 91.2\% average of \ours. SVEN is better in two individual settings, Gemma3-4B on C\texttt{++} and Gemma3-12B on Java, but \ours leads in the other ten without updating any model parameters. On Gemma3-4B/C++, the target model appears sensitive to detailed security context: the cue-only Top-3 variant reaches 86.6\% (in~\autoref{tab:cybernative-cwe-cue-only-topk}), substantially above the 57.0\% full-context result. SVEN also benefits from model- and language-specific training on the CyberNative split. On Gemma3-12B/Java, the difference is only 1.4 points, corresponding to two of 142 samples, and does not indicate a systematic disadvantage.

In addition, the CyberNative benchmark is demanding because some coding questions explicitly request unsafe implementation choices. A task may, for example, instruct the model to copy data using \texttt{strcpy}, even when the destination buffer is too small. A model that follows this instruction literally produces a known memory-safety vulnerability. A code assistant should not assume that the user understands the security consequences of a requested API. It should preserve the intended program behavior while replacing dangerous implementation details with safe alternatives. The results indicate that a generic request for secure code is often insufficient for this setting, whereas task-specific CWE analysis gives the target model enough context to recognize and avoid the relevant weakness.
The gains are also stable across five independently sampled evaluation splits with five random seeds, with language-level standard deviations between 0.92 and 1.77 percentage points (\autoref{sec:stability}).


\begin{table*}[t]
\centering
\caption{Safe code rate (\%) of commercial LLMs on CyberNative under
three prompting conditions.
\textit{Base} uses only the original task description.
\textit{Secure Prompt} explicitly requests secure code generation.
\textit{\ours} uses the black-box two-stage pipeline with LLM-selected,
task-relevant CWE guidance.
Provider refusals and invalid generations are counted as unsafe.}
\label{tab:api_inference}
\setlength{\tabcolsep}{4pt}
\begin{tabular}{@{}l ccc ccc ccc@{}}
\toprule
 & \multicolumn{3}{c}{C++}
 & \multicolumn{3}{c}{Java}
 & \multicolumn{3}{c}{Python} \\
\cmidrule(lr){2-4}
\cmidrule(lr){5-7}
\cmidrule(lr){8-10}
Model
  & Base & Secure Prompt & \ours
  & Base & Secure Prompt & \ours
  & Base & Secure Prompt & \ours \\
\midrule
Claude Haiku 4.5
  & 9.86  & 95.07 & \textbf{99.30}
  & 48.59 & 85.92 & \textbf{95.07}
  & 33.10 & 91.55 & \textbf{97.18} \\

Claude Sonnet 5
  & 6.34  & 95.07 & \textbf{100.00}
  & 40.85 & 89.44 & \textbf{97.18}
  & 14.08 & 96.48 & \textbf{98.59} \\
\midrule

DeepSeek V4 Flash
  & 7.04  & 95.07 & \textbf{99.30}
  & 34.51 & 85.21 & \textbf{93.66}
  & 9.86  & 74.65 & \textbf{100.00} \\

DeepSeek V4 Pro
  & 7.75  & 78.87 & \textbf{83.10}
  & 33.80 & 76.06 & \textbf{91.55}
  & 10.56 & 58.45 & \textbf{97.18} \\
\midrule

GPT-5.6 Luna
  & 7.75  & 88.73 & \textbf{95.77}
  & 40.85 & 85.21 & \textbf{90.85}
  & 15.49 & 71.83 & \textbf{97.18} \\

GPT-5.6 Sol
  & 10.56 & 84.51 & \textbf{95.77}
  & 40.85 & 83.10 & \textbf{90.14}
  & 20.42 & 82.39 & \textbf{95.77} \\
\bottomrule
\end{tabular}
\end{table*}

\begin{figure*}
    \centering
    \setlength{\abovecaptionskip}{-4pt}
    \includegraphics[width=0.9\linewidth]{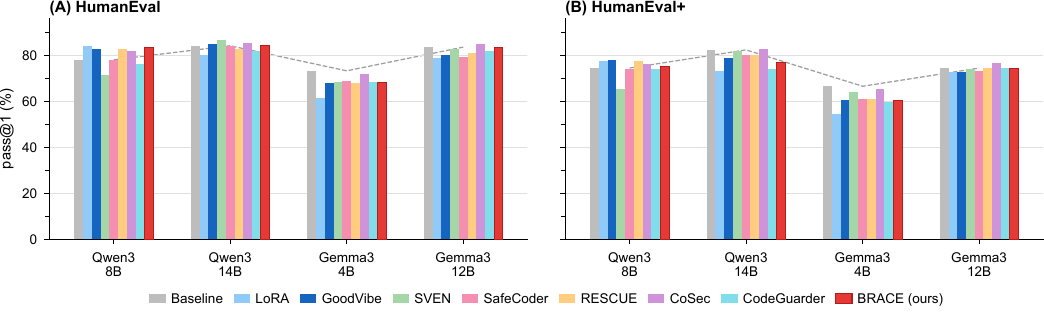}
    \caption{HumanEval pass@1 (\%) for 4 open-source models. (A) HumanEval base~\cite{chen2021codex}. (B) HumanEval+~\cite{liu2023evalplus}, with an expanded automatically generated test set.}
    \label{fig:humaneval}
\end{figure*}
  
\subsection{Black-Box Evaluation}
\label{sec:blackbox}

\autoref{tab:api_inference} evaluates six commercial models from Anthropic, DeepSeek, and OpenAI through their public text-generation interfaces. The target models are treated as black boxes: \ours{} does not access their parameters, hidden states, gradients, or output logits.

Recent frontier models still generate unsafe code when security is not stated explicitly. Under the Base condition, their average Safe Code Rate is only 21.8\%, despite their strong general coding ability. The problem is most severe for C\texttt{++}, where the average falls to 8.2\%. These failures should not be read as an inability to complete the coding task. A model may produce code that appears functionally plausible while following an unsafe API choice or omitting a required boundary check. Stronger code-generation capability therefore does not imply that security constraints will be inferred automatically.

Secure Prompt raises the average Safe Code Rate substantially, which suggests that these models already contain useful secure-coding knowledge. Its effect, however, remains uneven across providers and languages. A general request for secure code does not tell the model which weakness is relevant or which part of the implementation requires attention. The remaining failures are concentrated in settings where the task admits a straightforward but unsafe solution, particularly memory-related C\texttt{++} tasks and several Python tasks handled by the DeepSeek and GPT models.

Adding task-specific CWE guidance by \ours raises the average Safe Code Rate from 84.3\% under Secure Prompt to 95.4\%. The improvement occurs in all 18 model--language settings and is largest where the generic security instruction is least effective. This pattern supports the distinction between possessing security knowledge and applying it at the right point in generation. The frontier models often respond to a general security request, but the risk analysis makes that response more consistent by identifying the concrete weakness that should guide the implementation.

The gains also transfer across model families without provider-specific tuning. Refusals and invalid generations are counted as unsafe, so the result cannot be explained by models avoiding difficult tasks. Instead, \ours{} improves the rate at which valid generations account for the relevant security risk while preserving the ordinary black-box deployment interface. The results show that unsafe generation remains a practical problem even for current frontier models, and that model scale alone does not remove the need for explicit, task-level security guidance.

\subsection{Utility Preservation}
\label{sec:utility}
We evaluate whether security guidance reduces the functional correctness of generated code on HumanEval and HumanEval+. Figure~\ref{fig:humaneval} reports pass@1 for four models and nine generation methods.

On HumanEval, \ours preserves the utility of the underlying models. Its average pass@1 is 79.9\%, compared with 79.7\% for the unguided baseline. \ours matches or improves upon the baseline for three of the four models, including a gain of 5.5 percentage points on Qwen3-8B. The only noticeable decrease occurs on Gemma3-4B, where pass@1 drops by 4.9 points; one possible explanation is that the small Gemma3-4B is more sensitive to additional constraints.
Averaged across the four models, \ours ranks second among the nine methods, behind CoSec. These results indicate that adding task-specific security knowledge generally does not distract the model from the original programming objective. HumanEval+ shows a broadly similar trend, with \ours remaining competitive while incurring only a modest average reduction relative to the unguided baseline.

Overall, \ours does not introduce a substantial utility penalty. Its performance remains close to the baseline and competitive with methods that require model fine-tuning or specialized decoding. The results suggest that the security context usually complements the task description rather than replacing it, allowing \ours to improve secure code generation without materially compromising functional correctness.

\begin{table*}[t]
  \centering
  \caption{CWEval pass@10 results (\%) across six models. CWEval generates 20 samples per coding problem by default; we generate 10 samples per problem to reduce evaluation costs. F+S requires both functionality and security to pass.}
  \label{tab:cweval-pass10-all-methods}
  \setlength{\tabcolsep}{3.5pt}
    \begin{tabular}{l*{15}{c}}
      \toprule
      & \multicolumn{3}{c}{Base}
      & \multicolumn{3}{c}{Secure Prompt}
      & \multicolumn{3}{c}{RESCUE~\cite{shi2026rescue}}
      & \multicolumn{3}{c}{CodeGuarder~\cite{lin2025codeguarder}}
      & \multicolumn{3}{c}{\ours} \\
      \cmidrule(lr){2-4}\cmidrule(lr){5-7}\cmidrule(lr){8-10}\cmidrule(lr){11-13}\cmidrule(lr){14-16}
      Model
      & Func. & Sec. & F+S
      & Func. & Sec. & F+S
      & Func. & Sec. & F+S
      & Func. & Sec. & F+S
      & Func. & Sec. & F+S\\
      \midrule
      Qwen3-8B
        & 78.99 & 50.42 & 43.70
        & \textbf{80.67} & 63.87 & 57.14
        & \textbf{80.67} & \textbf{76.47} & \textbf{65.55}
        & 72.27 & 63.87 & 53.78
        & 77.31 & 75.63 & \textbf{65.55} \\
      Qwen3-14B
        & \textbf{86.55} & 61.34 & 52.94
        & 85.71 & 74.79 & 68.91
        & 82.35 & 76.47 & 69.75
        & 81.51 & 72.27 & 63.87
        & 85.71 & \textbf{81.51} & \textbf{78.15} \\
    Qwen3.6-27B
        & \textbf{93.28} & 71.43 & 71.43
        & \textbf{93.28} & 87.39 & 87.39
        & 90.76 & 88.24 & 87.39
        & \textbf{93.28} & 86.55 & 86.55
        & 91.60 & \textbf{89.08} & \textbf{89.08} \\
      Gemma3-4B
        & \textbf{62.18} & 38.66 & 30.25
        & 57.98 & 37.82 & 30.25
        & 57.98 & 42.02 & 31.09
        & 53.78 & 39.50 & 27.73
        & 59.66 & \textbf{52.94} & \textbf{42.86} \\
      Gemma3-12B
        & 75.63 & 42.86 & 39.50
        & \textbf{76.47} & 53.78 & 48.74
        & 68.07 & 52.10 & 47.90
        & 68.07 & 45.38 & 38.66
        & 72.27 & \textbf{68.07} & \textbf{59.66} \\
      Gemma4-31B
        & 89.92 & 61.34 & 60.50
        & \textbf{90.76} & 80.67 & 78.99
        & \textbf{90.76} & \textbf{87.39} & \textbf{86.55}
        & \textbf{90.76} & 80.67 & 78.15
        & 86.55 & \textbf{87.39} & 82.35 \\
      \midrule
      Average
        & \textbf{81.09} & 54.34 & 49.72
        & 80.81 & 66.39 & 61.90
        & 78.43 & 70.45 & 64.71
        & 76.61 & 64.71 & 58.12
        & 78.85 & \textbf{75.77} & \textbf{69.61} \\
      \bottomrule
    \end{tabular}%
\end{table*}

\subsection{Joint Security and Functionality-CWEval}
\label{sec:cweval_results}

CWEval does not provide the paired training split needed to reproduce the benchmark-specific adaptation used by several fine-tuning baselines in our CyberNative evaluation. We therefore restrict this comparison to methods that can be applied uniformly without adapting the target model: Base, Secure Prompt, RESCUE, CodeGuarder, and \ours.

\autoref{tab:cweval-pass10-all-methods} reports the results on all 119 CWEval tasks. Func and Sec indicate whether the sampled set contains a completion that passes the functionality or security tests, respectively. F+S is stricter because the same completion must pass both. This distinction matters: a method may produce one functionally correct completion and a different secure completion without ever producing code that satisfies both requirements.

The Base results expose security as the main constraint. The models solve a large fraction of the functional tests, but many of these solutions fail the corresponding security checks, leaving an average F+S rate below 50\%. Secure Prompt reduces this gap without materially changing functionality, confirming that an explicit security instruction is useful. Its remaining failures show that a general instruction often does not identify the implementation decision responsible for the weakness. Retrieval-based guidance improves security further, but the additional context can also distract the model from the functional specification. This trade-off is most visible for CodeGuarder, whose lower functionality offsets much of its security gain.

\ours achieves the highest average security and F+S scores while keeping functionality close to the unguided model. Its average F+S rate is 69.6\%, an improvement of 19.9 percentage points over Base and 4.9 points over RESCUE, the strongest competing method in this experiment. It obtains the best or tied-best F+S result on five of the six models. The exception is Gemma4-31B, where \ours and RESCUE have the same security score, but RESCUE retains slightly higher functionality. This case illustrates why the joint metric is needed: stronger security alone does not guarantee a better usable solution.

The CWEval results complement the separate CyberNative and HumanEval evaluations. CyberNative shows that \ours increases the rate of secure generations, while HumanEval measures whether general coding ability is preserved on ordinary programming tasks. CWEval tests both properties on the same problem and the same completion. The improvement in F+S therefore provides direct evidence that task-specific risk guidance helps models generate code that is not merely safer in isolation, but remains executable and functionally correct.
The wall-clock generation time on CWEval is in~\autoref{sec:time_con}.

\begin{table}[t]
  \centering
  \small
  \caption{BaxBench pass@1 results (\%) for Gemma4-31B.}
  \label{tab:baxbench-gemma31b-stage2k1-by-language}
  \begin{tabular}{llrrr}
    \toprule
    Language & Method & Function & Security & F+S \\
    \midrule
    \multirow{3}{*}{\shortstack{Python}}
      & Base          & \textbf{73.21} & 38.39          & 36.61 \\
      & Secure Prompt & 70.54          & 43.75          & 41.96 \\
      & BRACE         & 66.96          & \textbf{48.21} & \textbf{43.75} \\
    \midrule
    \multirow{3}{*}{\shortstack{JavaScript}}
      & Base          & \textbf{59.82} & 40.18          & 34.82 \\
      & Secure Prompt & 54.46          & 44.64          & 34.82 \\
      & BRACE         & 51.79          & \textbf{47.32} & \textbf{35.71} \\
    \midrule
    \multirow{3}{*}{\shortstack{Go}}
      & Base          & 64.29          & 39.29          & 28.57 \\
      & Secure Prompt & \textbf{70.24} & 46.43          & 34.52 \\
      & BRACE         & 69.05          & \textbf{58.33} & \textbf{48.81} \\
    \midrule
    \multirow{3}{*}{\shortstack{Overall}}
      & Base          & \textbf{65.91} & 39.29          & 33.77 \\
      & Secure Prompt & 64.61          & 44.81          & 37.34 \\
      & BRACE         & 62.01          & \textbf{50.65} & \textbf{42.21} \\
    \bottomrule
  \end{tabular}
\end{table}

\subsection{Project-level Evaluation by BaxBench}
\label{sec:baxbench}
Unlike function-level benchmarks, BaxBench~\cite{vero2025baxbench} evaluates complete project-level backend applications. Generated projects are built and launched inside isolated sandboxes, after which the evaluator sends real requests to their API endpoints and exercises database, filesystem, multimedia, and framework-dependent behavior. Given the difficulty of these generation tasks, we restrict our BaxBench evaluation to a relatively large model as shown in~\autoref{tab:baxbench-gemma31b-stage2k1-by-language}. BRACE consistently provides the strongest balance between functionality and security across all three languages, although this improvement is accompanied by a modest reduction in functional correctness. The gain is particularly pronounced for Go, suggesting that targeted risk-conditional generation is more effective than instruction-only security guidance when vulnerabilities can be addressed without substantially altering program structure. Since BaxBench executes complete projects in a sandbox and exercises their backend services, these results demonstrate that BRACE’s security improvements generally survive deployment-like execution rather than merely producing code that appears secure statically.

\begin{table}[t]
\centering
\caption{Safe Code Rate (\%) of \ours{} and Random-CWE guidance on CyberNative. $\Delta$ denotes \ours{} minus Random.}
\label{tab:brace_random}
\small
\begin{tabular}{llccc}
\toprule
Model & Language & \ours{} & Random & $\Delta$ \\
\midrule
\multirow{3}{*}{Qwen3-8B}
& C\texttt{++} & 94.4 & 14.1 & $80.3$ \\
& Java          & 95.8 & 66.9 & $28.9$ \\
& Python        & 90.8 & 40.1 & $50.7$ \\
\midrule

\multirow{3}{*}{Qwen3-14B}
& C\texttt{++} & 97.9 & 16.2 & $81.7$ \\
& Java          & 96.5 & 72.5 & $24.0$ \\
& Python        & 98.6 & 43.0 & $55.6$ \\
\midrule

\multirow{3}{*}{Gemma-3-4B}
& C\texttt{++} & 57.0 & 14.1 & $42.9$ \\
& Java          & 90.1 & 50.0 & $40.1$ \\
& Python        & 90.8 & 28.9 & $61.9$ \\
\midrule

\multirow{3}{*}{Gemma-3-12B}
& C\texttt{++} & 97.2 & 87.3 &  $9.9$ \\
& Java          & 90.1 & 52.1 & $38.0$ \\
& Python        & 95.1 & 45.1 & $50.0$ \\
\bottomrule

\end{tabular}
\end{table}

\begin{figure}[t]
    \centering
    \setlength{\abovecaptionskip}{-4pt}
    \includegraphics[width=\linewidth]{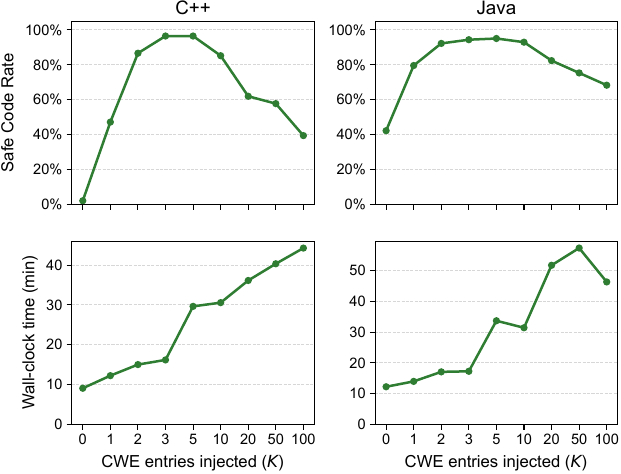}
    \caption{Effect of CWE retrieval depth on safe code generation. Safe Code Rate (top) and wall-clock inference time (bottom) as a function of the number of CWE entries injected into the prompt ($K$), evaluated across C++ and Java.}
    \label{fig:cweablation}
\end{figure}

\subsection{Ablation Study}
\label{sec:ablation}

\noindent
\textbf{Knowledge activation rather than injection.}
Table~\ref{tab:cybernative-cwe-cue-only-topk} shows that a compact risk cue is often sufficient to elicit secure behavior. Qwen3-14B and Gemma-3-12B achieve consistently high safety rates across all three languages, even though the prompt contains only CWE identifiers and canonical names. No vulnerability descriptions, mitigations, or secure-coding recipes are provided. The improvement therefore cannot be attributed to injecting detailed security knowledge into the context. Instead, the CWE cue serves as a task-specific key that activates knowledge already encoded in the target model.

Using more than one CWE generally improves safety, particularly for models and languages with weaker Top-1 results. The trend is not universal, however. A few model--language pairs perform slightly better with fewer cues, suggesting that an irrelevant CWE can dilute the intended signal. The effectiveness of activation thus depends more on selecting the right risk concepts than on increasing the amount of security text. BRACE should be viewed as a mechanism for identifying a small set of relevant knowledge anchors, rather than a mechanism for copying external security knowledge into the prompt. This distinction also explains the limitation of a generic secure prompt instruction: it states the desired outcome but does not tell the model which part of its security knowledge is relevant to the current task.
\begin{table}[t]
  \centering
  \small
  \setlength{\tabcolsep}{5.5pt}
  \caption{Safe Code Rate (\%) on CyberNative when Stage 2 receives only CWE IDs and canonical names. CWE descriptions and mitigations are excluded. Each language contains 142 held-out tasks. Bold indicates the best $K$ for each model$\times$language pair.}
  \label{tab:cybernative-cwe-cue-only-topk}
  \begin{tabular}{llccc}
    \toprule
    Model & Language & $K=1$ & $K=2$ & $K=3$ \\
    \midrule
    \multirow{3}{*}{\shortstack{Qwen3\\8B}}
      & C\texttt{++} & 68.3 & 72.5 & \textbf{73.2} \\
      & Java         & 82.4 & 89.4 & \textbf{90.1} \\
      & Python       & 66.9 & 73.9 & \textbf{76.8} \\
    \midrule
    \multirow{3}{*}{\shortstack{Qwen3\\14B}}
      & C\texttt{++} & 96.5 & 95.1 & \textbf{97.2} \\
      & Java         & 90.8 & 96.5 & \textbf{97.9} \\
      & Python       & \textbf{96.5} & 94.4 & 95.8 \\
    \midrule
    \multirow{3}{*}{\shortstack{Gemma-3\\4B}}
      & C\texttt{++} & \textbf{87.3} & 85.9 & 86.6 \\
      & Java         & 78.9 & 80.3 & \textbf{81.0} \\
      & Python       & 64.1 & 74.6 & \textbf{75.4} \\
    \midrule
    \multirow{3}{*}{\shortstack{Gemma-3\\12B}}
      & C\texttt{++} & 94.4 & \textbf{95.8} & \textbf{95.8} \\
      & Java         & 84.5 & 88.0 & \textbf{90.1} \\
      & Python       & 85.2 & 95.1 & \textbf{96.5} \\
    \bottomrule
  \end{tabular}
\end{table}

\noindent
\textbf{Random CWE.}
We examine whether \ours benefits from identifying task-relevant weaknesses or merely from adding CWE-related security text to the prompt. We keep the target model, the number of CWE entries, the canonical grounding format, and the decoding settings fixed, but replace the selector's predictions with valid CWE identifiers sampled at random from the same catalog.
Across the 12 model$\times$language settings in \autoref{tab:brace_random}, \ours achieves an average Safe Code Rate of 91.2\%, compared with 44.2\% under Random-CWE guidance, a difference of 47.0 percentage points. \ours performs better in every setting. Random guidance occasionally works well when a sampled weakness happens to be relevant, but its effectiveness varies considerably across models and languages. By contrast, \ours remains consistently strong across both model families and all three languages.

\begin{figure*}[t]
    \centering
    \setlength{\abovecaptionskip}{-4pt}
    \includegraphics[width=1\textwidth]{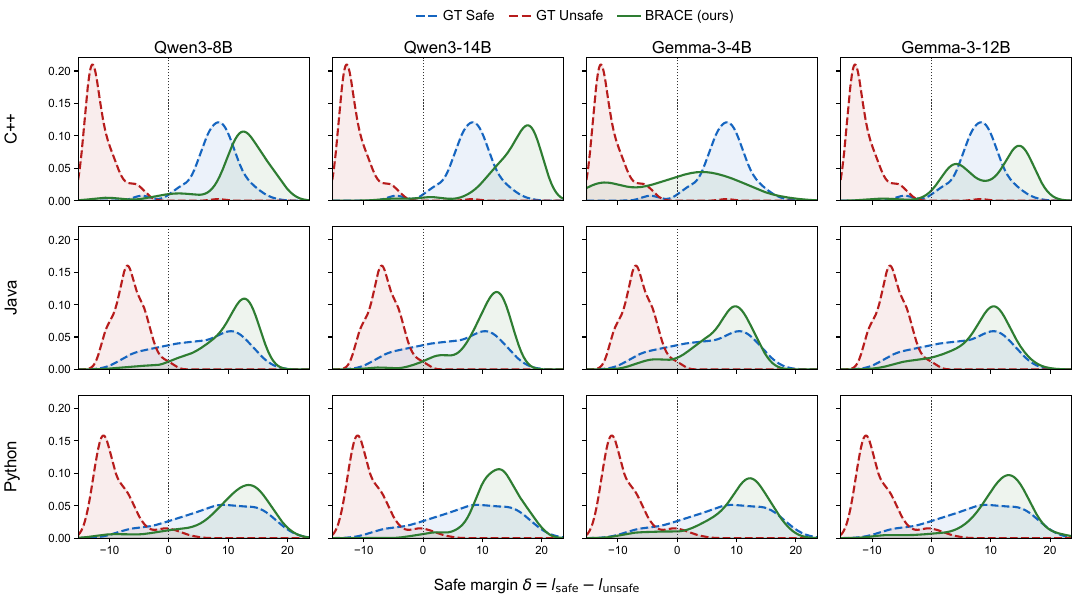}
    \caption{Kernel density estimates of the judge safe margin for secure reference code, insecure reference code, and code generated by \ours. GT refers to ground truth. The dotted vertical line marks the decision boundary $\delta=0$. All panels use the same axis limits and contain 142 samples from each group.}
    \label{fig:composite_logit_kde}
\end{figure*}

Since the two conditions use the same CWE catalog and downstream prompt construction, the difference comes from which weaknesses are selected rather than from the presence of security terminology alone. Canonical guidance is useful only when it addresses risks that are plausible for the current task. This result confirms that the selector is a necessary part of \ours: it identifies the security knowledge that the target model needs during generation.

\noindent
\textbf{Retrieval Depth $K$.}
\autoref{fig:cweablation} shows how the safe code rate and wall-clock time vary with the number of CWE entries ($K$), evaluated on C\texttt{++} and Java. Both languages follow an inverted U of safe code rate: the rate climbs steeply at small $K$, rising from near zero to its peak around $K\in\{3,5\}$, and then falls monotonically as $K$ grows further, dropping back below 40\% on C\texttt{++} by $K{=}100$. Java exhibits the same shape with a gentler decline. This inversion is consistent with an attention-dilution effect: once the injected context grows large, the security guidance relevant to the query is buried among many redundant entries, and the model can no longer resolve the specific vulnerability it faces. A moderate retrieval depth of $K\in\{3,5\}$ therefore offers the best balance between knowledge coverage and prompt coherence.

Increasing $K$ also carries a substantial latency cost. On the C\texttt{++} evaluation set, generation time grows roughly 5-fold between $K{=}0$ and $K{=}100$, with a comparable trend on Java. The growth is a direct consequence of the enlarged prompt: each additional CWE entry appends description and mitigation text to every query, expanding the input processed during the prefill phase and lengthening per-sample decoding. 

\subsection{Judge-Model Representation Analysis}
\label{sec:judge_representation}

The Safe Code Rate reports the final decision of the judge but does not show how confidently the generated code is separated from unsafe code. We therefore use the fixed judge classifier as a post-hoc feature extractor. We analyze Qwen3-8B, Qwen3-14B, Gemma-3-4B, and Gemma-3-12B on C\texttt{++}, Java, and Python. Each setting contains three equally sized groups: code generated by \ours, held-out secure reference code, and the corresponding insecure reference code. All groups are evaluated with the same 512-token limit used in the main security evaluation. Given a code snippet $x$, we extract the last-layer representation of its final non-padding token, denoted by $h(x)$. The classification head produces the safe and unsafe logits as
\begin{equation}
    \begin{bmatrix}
        l_{\mathrm{safe}}(x) \\
        l_{\mathrm{unsafe}}(x)
    \end{bmatrix}
    =
    W_{\mathrm{cls}}h(x)+b_{\mathrm{cls}}.
\end{equation}
We summarize the judge's decision using the safe margin
\begin{equation}
    \delta(x)
    =
    l_{\mathrm{safe}}(x)-l_{\mathrm{unsafe}}(x).
\end{equation}
A positive margin corresponds to a safe prediction, while a negative margin corresponds to an unsafe prediction.

\autoref{fig:composite_logit_kde} shows that the secure and insecure reference distributions lie on opposite sides of the decision boundary in all three languages. The outputs of \ours generally follow the secure side of this separation. Across the 12 model$\times$language settings, their average distributional overlap with secure reference code is 0.63, compared with 0.09 for insecure reference code. The overlap with the secure distribution is larger in every setting. The effect therefore extends beyond moving a small number of borderline samples across the decision boundary: most generated outputs share features that the judge associates with secure code.

The green and blue curves do not always coincide. This is expected because generated programs can differ from the reference implementations in structure, API choice, and coding style while remaining secure. Qwen3-14B, for example, often receives margins beyond the mode of the secure reference distribution. Such a shift indicates greater classifier confidence rather than a claim that the generated code is intrinsically more secure than the reference code.

Gemma-3-4B on C\texttt{++} is the main exception. Its margin distribution is broad and crosses the decision boundary, unlike the more concentrated positive distributions observed in the other settings. This pattern agrees with its weaker C\texttt{++} result in the main evaluation. The figure thus identifies a limitation that is hidden by the aggregate result: security guidance remains less reliable when the target model cannot consistently translate the supplied controls into a safe implementation.

\begin{figure}
    \centering
    \setlength{\abovecaptionskip}{-4pt}
    \includegraphics[width=1\linewidth]{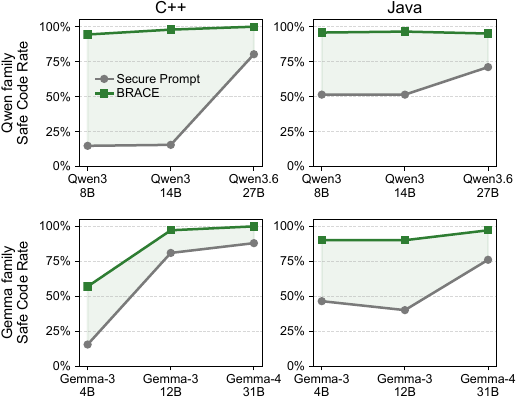}
    \caption{Safe Code Rate (\%) on CyberNative across the Qwen and Gemma model families in C\texttt{++} and Java.}
    \label{fig:scaling}
\end{figure}

These observations support a limited behavioral conclusion. Task-specific CWE guidance changes the generated programs in ways that the fixed judge consistently associates with secure code. The analysis does not establish that \ours modifies a particular internal direction in the target model, nor does it identify a causal representation inside that model. It is instead consistent with the intended role of the harness: making relevant secure-coding behavior easier to express through contextual guidance, without fine-tuning or access to model internals.

\subsection{Model Architecture Scaling}
\label{sec:modelscaling}

We examine how the target model affects the use of task-specific security guidance. The Qwen series contains Qwen3-8B, Qwen3-14B, and Qwen3.6-27B, while the Gemma series contains Gemma-3-4B, Gemma-3-12B, and Gemma-4-31B. For a given task, all target models receive the same cached risk predictions from the Gemma-4-31B selector and the same canonical CWE context. The comparison therefore changes the target generator while holding the security analysis fixed.

\autoref{fig:scaling} shows that a larger target model does not produce a smooth improvement under Secure Prompt. Qwen3-8B and Qwen3-14B perform similarly in both C\texttt{++} and Java despite their difference in size. The improvement becomes clear only with Qwen3.6-27B. The Gemma series is also uneven: Gemma-3-12B performs well on C\texttt{++} but remains much weaker on Java. Model capacity helps, but its effect depends on both the model architecture and the programming language.

\ours{} produces a different scaling pattern. Across the 12 model--language settings shown in the figure, it raises the macro-average Safe Code Rate from 52.6\% to 92.6\% and improves every individual setting. Qwen3-8B already exceeds 94\% in both languages, indicating that the method does not require a large target model to be effective. The two largest models obtain rates above 95\% throughout, suggesting that greater capacity mainly improves the consistency with which the supplied security controls are implemented.

\section{Conclusion}
\label{sec:conclusion}
This work studies secure code generation as a problem of risk-conditioned knowledge activation. We introduce \ours, a training-free, black-box harness that uses an LLM to identify task-relevant CWE risks, validates them against the official CWE catalog, and supplies concise canonical guidance to the target model. The method requires no fine-tuning, example-based code knowledge base, or access to model parameters, hidden states, and logits. Its risk analysis can also be cached and reused across target models.

On CyberNative, \ours raises the average Safe Code Rate from 33.9\% under Secure Prompt to 91.2\% on CyberNative on open-weight models and from 84.3\% to 95.4\% on commercial models. It also improves CWEval Func-Sec@10 from 49.7\% to 69.6\% while preserving HumanEval pass@1. These findings suggest that modern code models already possess useful secure-coding knowledge, but often require task-specific risk guidance to apply it consistently.


\bibliographystyle{plain}
\bibliography{references}

\appendix

\section{Discussion: Limitations and Future Work}
\label{sec:discussion}

\noindent
\textbf{Limitations.} \ours performs risk analysis before code generation and therefore focuses on vulnerabilities that are foreseeable from the task specification. Implementation-specific weaknesses that emerge only from particular generation choices may not be anticipated by the selector. Combining task-level risk analysis with lightweight post-generation verification is a promising direction for future work.

\noindent
\textbf{Future Work.} Future work could explore a closed-loop extension that complements task-level risk localization with lightweight post-generation implementation analysis, allowing newly emerged risks to trigger targeted mitigation or regeneration.

\section{Judge Reliability on Generated Code}
\label{sec:judge_generated}

Our security metric depends on the LLM judge classifying model-generated code, which differs in distribution from the training data. To verify the judge on this actual evaluation target, we manually annotate 142 generated Python samples and compare the judge against two static analyzers (\autoref{tab:judge_on_generated}). This annotation is cross-validated with LLM models, including Gemini 3.5 Flash, GPT-5.3-mini, and Claude Opus 4.8. 
The judge attains 0.838 accuracy, far above CodeQL (0.430) and ICD (0.606), confirming that it remains the most reliable detector on generated code. As expected, this is lower than its 0.955 accuracy on the in-distribution dataset (\autoref{tab:static_vs_llm}), reflecting the harder out-of-distribution setting; the judge nonetheless preserves a large and consistent margin over static tools, supporting its use as the primary evaluator.

\begin{table}[h]
\centering
\caption{Detection performance on generated Python code (142 samples, human-annotated) for static analysis tools versus the LLM judge. This validates the judge on actual model-generated outputs rather than the original dataset.}
\label{tab:judge_on_generated}
\setlength{\tabcolsep}{3pt}
\begin{tabular}{lccccccc}
\toprule
\multirow{2}{*}{\textbf{Method}} & \multicolumn{4}{c}{\textbf{Confusion Matrix}} & \multicolumn{3}{c}{\textbf{Key Metrics}}\\
\cmidrule(lr){2-5} \cmidrule(lr){6-8}
& \textbf{TP} & \textbf{FP} & \textbf{TN} & \textbf{FN} & \textbf{Precision} & \textbf{Recall} & \textbf{Acc.} \\
\midrule
\rowcolor[gray]{.95} \multicolumn{8}{l}{\textit{Static Tools}} \\
CodeQL & 20 & 41 & 41 & 40 & 0.328 & 0.333 & 0.430 \\
ICD    & 41 & 37 & 45 & 19 & 0.526 & 0.683 & 0.606 \\
\midrule
\rowcolor[gray]{.95} \multicolumn{8}{l}{\textit{LLM Judge}} \\
\textbf{Qwen3-0.6b} & \textbf{47} & \textbf{10} & \textbf{72} & \textbf{13} & \textbf{0.825} & \textbf{0.783} & \textbf{0.838} \\
\bottomrule
\end{tabular}
\end{table}

\begin{table*}
\centering
\small
\setlength{\tabcolsep}{7pt}
\renewcommand{\arraystretch}{1.08}
\caption{Wall-clock generation time on CWEval. Each condition covers all
119 tasks with ten samples per task and a generation batch size of 10.
Times are in seconds and rounded to the nearest integer; values in
parentheses are relative to Secure Prompt for the same target model.
The measurement includes per-task retrieval and all auxiliary generation
calls, but excludes judging and sandbox execution.}
\label{tab:gen_time}
\begin{tabular}{lccccc}
\toprule
\multirow{2}{*}{\textbf{Model}}
& \multirow{2}{*}{\textbf{Secure Prompt}}
& \multirow{2}{*}{\textbf{RESCUE}}
& \multirow{2}{*}{\textbf{CodeGuarder}}
& \multicolumn{2}{c}{\textbf{BRACE}} \\
\cmidrule(lr){5-6}
& & & & \textbf{Cached} & \textbf{Uncached} \\
\midrule
Qwen3-8B
& 1,365
& 4,309 (\(3.16\times\))
& 1,693 (\(1.24\times\))
& 1,994 (\(1.46\times\))
& 2,766 (\(2.03\times\)) \\

Qwen3-14B
& 1,475
& 5,338 (\(3.62\times\))
& 2,016 (\(1.37\times\))
& 2,181 (\(1.48\times\))
& 2,953 (\(2.00\times\)) \\

Gemma-3-4B
& 1,620
& 4,492 (\(2.77\times\))
& 2,115 (\(1.31\times\))
& 2,716 (\(1.68\times\))
& 3,488 (\(2.15\times\)) \\

Gemma-3-12B
& 3,770
& 10,928 (\(2.90\times\))
& 5,080 (\(1.35\times\))
& 7,041 (\(1.87\times\))
& 7,813 (\(2.07\times\)) \\
\bottomrule
\end{tabular}
\end{table*}

\section{Experimental details} 
\label{sec:experimental_details}
We use random seed 42 for all experiments. For CyberNative, we generate one completion per task with temperature 0 and top-$p=1.0$, using a maximum of 600 new tokens for local models and 1,000 for API models. For HumanEval, we generate one completion with temperature 0.2, top-$p=0.95$, and a 512-token limit. For CWEval, we sample 10 completions per task with temperature 0.8, top-$p=1.0$, and a 2,048-token limit. BRACE's Stage-1 selector uses deterministic decoding. Local inference is performed with one NVIDIA A100-SXM4-80GB GPU per worker. For API models, optional thinking is disabled (\texttt{reasoning\_effort=none} or the provider-equivalent setting). CyberNative outputs are classified by the fine-tuned Qwen3-0.6B judge.

\section{Baseline Implementation Details}
\label{sec:baseline_details}

We group the seven baselines in \autoref{tab:main} into fine-tuning methods, which update or attach trainable parameters, and retrieval/decoding methods, which act only at inference time. Unless noted otherwise, fine-tuning uses the CyberNative DPO training split (Section~\ref{sec:exp}) and generation uses up to 600 new tokens. Hyperparameters were selected based on prior studies~\cite{maximilian2024goodvibe,hu2022lora} or standard practice in the field.

\textbf{LoRA}~\cite{hu2022lora} applies standard low-rank adaptation to the secure code of the DPO split, with adapters on \texttt{q/k/v/o\_proj} and \texttt{up/gate\_proj} ($r{=}8$, $\alpha{=}4$, dropout $0.05$), trained for 2 epochs at learning rate $2{\times}10^{-4}$ with 200 warmup steps, batch size 1, and maximum length 512.

\textbf{SafeCoder}~\cite{tony2023safecoder} performs instruction tuning on SafeCoder's own corpus (vulnerability-fixing commits plus general instruction-following data) with a security-aware objective: cross-entropy on the functional and corrected-code tokens and an uncertainty loss that penalizes high probability on insecure tokens. We implement it as LoRA on \texttt{q\_proj} and \texttt{v\_proj} ($r{=}8$, $\alpha{=}16$), trained for 3 epochs at learning rate $5{\times}10^{-5}$, batch size 4, maximum length 1024, loss weight $0.5$, with the insecure-token probability clamped at $0.9$ for numerical stability.

\textbf{GoodVibe}~\cite{maximilian2024goodvibe} performs neuron-selective fine-tuning: safety-relevant neurons are identified by a logistic probe (z-score $>3$ and positive weight), grouped with $k$-means clustering, and only the clustered neurons are updated through a tied-parameter mask. We use the top-50 safety neurons per layer, 2 epochs, learning rate $1{\times}10^{-4}$, and batch size 1.

\textbf{SVEN}~\cite{he2023sven} learns a secure-code prefix prepended as virtual tokens at inference while the backbone stays frozen. The number of virtual tokens scales with model size (5 for 4B, 8 for 7--8B, and 12 for 12--14B models); trained for 5 epochs at learning rate $1{\times}10^{-2}$, batch size 4, and maximum length 1024.

\textbf{RESCUE}~\cite{shi2026rescue} uses hierarchical multi-faceted retrieval: the model first drafts a solution, a vulnerability-cause analysis is produced (up to 150 tokens), and CWE-level and code-level entries are retrieved from the RESCUE knowledge base via reciprocal-rank fusion (thresholds $[4.0, 0.75]$ at the CWE level and $[4.0, 0.75, 0.65]$ at the code level, keeping the top-4 CWEs and top-1 code). The retrieved security guidelines and secure code example are appended to the prompt for a final regeneration. RESCUE provides no knowledge base for Java, which therefore falls back to zero-shot.

\textbf{CodeGuarder}~\cite{lin2025codeguarder} retrieves the top-3 root causes and their reference code from a large ReposVul-based vulnerability knowledge base by embedding similarity, and prepends a structured block of security knowledge and reference examples to the prompt before generation.

\textbf{CoSec}~\cite{li2024cosec} co-decodes with a smaller security expert model (CodeGen-350M with CoSec's released weights). At each step the base token is accepted only if $\text{threshold} < \min(1,\, P(t_b \mid \text{base}) / P(t_e \mid \text{expert}))$, steering the base model toward the expert's secure preferences; we use an acceptance threshold of $0.3$ and decoding temperature $0.4$.

\begin{table}[t]
\centering
\caption{Exact CWE selection accuracy on the 119 CWEval tasks. Exact@$k$ measures whether the ground-truth CWE ID appears among the first $k$ predictions. Taxonomic near-misses are counted as incorrect.}
\label{tab:stage1-cwe-selection}
\small
\setlength{\tabcolsep}{4pt}
\begin{tabular}{lcccc}
\toprule
\textbf{Selector Model} & \textbf{Exact@1} & \textbf{Exact@2} & \textbf{Exact@3} & \textbf{Parse Fail.} \\
\midrule
Qwen3-8B       & 12.61\% & 30.25\% & 32.77\% & 0  \\
Qwen3-14B      & 21.01\% & 27.73\% & 27.73\% & 0  \\
Gemma-3-4B     &  5.04\% &  6.72\% &  9.24\% & 35 \\
Gemma-3-12B    & 12.61\% & 18.49\% & 21.85\% & 0  \\
Qwen3.6-27B    & 32.77\% & 48.74\% & 48.74\% & 0  \\
Gemma-4-31B    & \textbf{45.38\%} & \textbf{53.78\%} & \textbf{57.98\%} & 0  \\
\bottomrule
\end{tabular}
\end{table}

\section{Time Consumption}
\label{sec:time_con}

\autoref{tab:gen_time} reports the inference cost on all 119 CWEval tasks, with ten samples per task and a generation batch size of 10. RESCUE~\cite{shi2026rescue} is the slowest method because it invokes the target model three times to produce a draft, analyze possible vulnerability causes, and generate the final solution. Its cost is therefore dominated by repeated decoding rather than retrieval itself. CodeGuarder~\cite{lin2025codeguarder} remains closer to Secure Prompt because it performs one retrieval step followed by a single target-model call. With Stage~1 predictions cached, \ours{} also requires only one target-model call, and its runtime is \(1.46\times\) to \(1.87\times\) that of Secure Prompt. This overhead comes from processing the task-specific CWE context and generating the resulting solution, rather than from iterative repair or token-level intervention. Running Stage~1 from scratch raises the total cost to approximately twice that of Secure Prompt, but this analysis is performed once per task rather than once per sample. Its output can be shared by all ten completions and reused across target models, making the cached condition representative of repeated sampling and cross-model evaluation. Although CodeGuarder is the fastest knowledge-guided method in this comparison, it depends on embedding retrieval over an external code knowledge base. In contrast, \ours{} uses only the canonical CWE catalog and a reusable task-level risk analysis, avoiding both a code-example knowledge base and the repeated target-model calls required by RESCUE.

\section{CWE selector}
\label{sec:cwe_selector}
Table~\ref{tab:stage1-cwe-selection} shows that selecting a precise CWE from a coding task remains difficult, even for larger models. Gemma-4-31B performs best, placing the reference CWE among its first three predictions for 57.98\% of the tasks. The gap between Exact@1 and Exact@3 suggests that the relevant weakness is often recognized but not ranked first, which supports passing a small set of candidate CWEs to Stage~2. The strict exact-match metric also hides some meaningful near-misses. Among the 50 tasks missed by Gemma-4-31B at Top-3, 13 contain a direct parent or child of the reference CWE. Counting these cases gives a diagnostic exact-or-direct-relative rate of 68.91\%. For example, all five language variants labeled as CWE-943 are predicted as CWE-89, its direct child. Such predictions capture the intended injection risk and may still provide useful guidance, but they are not treated as correct because related CWEs can require different controls. Broader relations are even less reliable: confusing SSRF with open redirect, for instance, identifies a related input-handling problem but does not provide sufficient mitigation for SSRF. We therefore retain Exact@3 as the primary metric and report direct parent/child near-misses only as a secondary diagnostic. We use Gemma-4-31B as a fixed Stage-1 selector, selected a priori as the largest open-weight model in our candidate pool. \autoref{tab:stage1-cwe-selection} is a post-hoc diagnostic analysis and was not used for model selection.

\begin{table}
  \centering
  \small
  \setlength{\tabcolsep}{2pt}
  \caption{Safe Code Rate (\%) on CyberNative, with 142 held-out tasks per language. Draft is the self-audit-and-repair's first-stage code; RK only and Audit+RK are its two second-stage conditions. Best result per model$\times$language row is shown in \textbf{bold}.}
  \label{tab:cybernative-circa-risk-reasoning}
  \begin{tabular}{llccccc}
    \toprule
    \multirow{2}{*}{Model} & \multirow{2}{*}{Lang.}
      & \multirow{2}{*}{Sec. Prompt} & \multicolumn{3}{c}{Self-Audit-and-Repair}
      & \multirow{2}{*}{BRACE} \\
    \cmidrule(lr){4-6}
    & & & Draft & RK & Audit+RK & \\
    \midrule
    \multirow{3}{*}{\shortstack{Qwen3\\8B}} & C\texttt{++} & 14.8 & 5.6 & 45.1 & 57.7 & \textbf{94.4} \\
     & Java & 51.4 & 27.5 & 44.4 & 49.3 & \textbf{95.8} \\
     & Python & 28.2 & 10.6 & 19.7 & 19.0 & \textbf{90.8} \\
    \midrule
    \multirow{3}{*}{\shortstack{Qwen3\\14B}} & C\texttt{++} & 15.5 & 3.5 & 68.3 & 69.7 & \textbf{97.9} \\
     & Java & 51.4 & 20.4 & 49.3 & 50.0 & \textbf{96.5} \\
     & Python & 23.9 & 14.8 & 49.3 & 60.6 & \textbf{98.6} \\
    \midrule
    \multirow{3}{*}{\shortstack{Gemma-3\\4B}} & C\texttt{++} & 15.5 & 9.9 & 57.7 & \textbf{71.1} & 57.0 \\
     & Java & 46.5 & 37.3 & 47.9 & 46.5 & \textbf{90.1} \\
     & Python & 20.4 & 25.4 & 40.8 & 46.5 & \textbf{90.8} \\
    \midrule
    \multirow{3}{*}{\shortstack{Gemma-3\\12B}} & C\texttt{++} & 81.0 & 7.0 & 83.1 & 85.2 & \textbf{97.2} \\
     & Java & 40.1 & 36.6 & 54.9 & 54.2 & \textbf{90.1} \\
     & Python & 17.6 & 15.5 & 46.5 & 48.6 & \textbf{95.1} \\
    \midrule
    \multicolumn{2}{l}{Average} & 33.9 & 17.8 & 50.6 & 54.9 & \textbf{91.2} \\
    \bottomrule
  \end{tabular}
\end{table}

\section{Self Risk Reasoning}
To test whether BRACE's gains can be reproduced by allowing the target model to inspect and repair its own output, we additionally consider a two-stage \textit{self-audit-and-repair} baseline implemented with context-isolated model calls. In Stage~1, the target model receives only the original coding task, generates a complete Draft, and then audits that exact Draft without revising it. The audit produces an Audit Key describing the suspected weakness and a Repair Key specifying the required control and the behavior that should be preserved. Stage~2 starts from a fresh context and receives the original task, the Draft, and the selected repair information. The Repair-Key-only condition exposes only the actionable control, whereas the Audit+Repair condition additionally provides the model's diagnosis. Both conditions reuse the same Stage~1 artifact, so their difference is restricted to the risk information made available during repair.

As shown in Table~\ref{tab:cybernative-circa-risk-reasoning}, the self-audit-and-repair baseline relies on the target model itself both to produce the Draft and to formulate a free-form diagnosis of what may be wrong with it. Risks that are not recognized during the audit cannot inform the subsequent repair, while vague or incomplete diagnoses may lead to controls that do not address the relevant implementation decision. Despite the improvement over Secure Prompt, the self-audit-and-repair baseline remains below BRACE in macro average and in eleven of the twelve model--language settings, with Audit+Repair outperforming BRACE in only one setting. BRACE instead decouples risk identification from target-code generation, validates the predicted weakness classes against a closed CWE catalog, and grounds them in canonical descriptions and mitigations before generation. These results suggest that explicit risk reasoning is useful but not sufficient by itself: more structured, validated, and task-relevant risk guidance enables the target model to apply secure-coding behavior more consistently.

\section{Stability}
\label{sec:stability}
We evaluate BRACE with Qwen3-8B and Gemma-3-12B on five independently sampled CyberNative held-out splits, each containing 142 tasks per language. Table~\ref{tab:cybernative-brace-multiseed} shows that BRACE remains effective across different task partitions. Gemma-3-12B achieves the higher average safety rate, while Qwen3-8B is more stable across seeds. Their contrasting language-level strengths suggest that BRACE reliably activates task-specific security knowledge, but its realization still depends on the target model's language capabilities. Since decoding is deterministic, the observed variance primarily reflects sensitivity to task composition rather than sampling noise.

\begin{table}[t]
  \centering
  \small
  \caption{Safe Code Rate (\%) of BRACE on CyberNative over five held-out split seeds. Each split contains 142 tasks per language. Results are reported as mean $\pm$ sample standard deviation.}
  \label{tab:cybernative-brace-multiseed}
  \begin{tabular}{lccc}
    \toprule
    Model & C\texttt{++} & Java & Python \\
    \midrule
    \shortstack{Qwen3 8B}
      & $92.68 \pm 1.77$
      & $96.06 \pm 1.18$
      & $92.11 \pm 0.92$ \\
    \shortstack{Gemma3 12B}
      & $95.35 \pm 1.46$
      & $91.97 \pm 1.70$
      & $96.48 \pm 1.32$ \\
    \bottomrule
  \end{tabular}
\end{table}

\section{Judge-Model Representation Analysis (t-SNE)}
\begin{figure*}
    \centering
    \setlength{\abovecaptionskip}{-4pt}
    \includegraphics[width=0.9\textwidth]{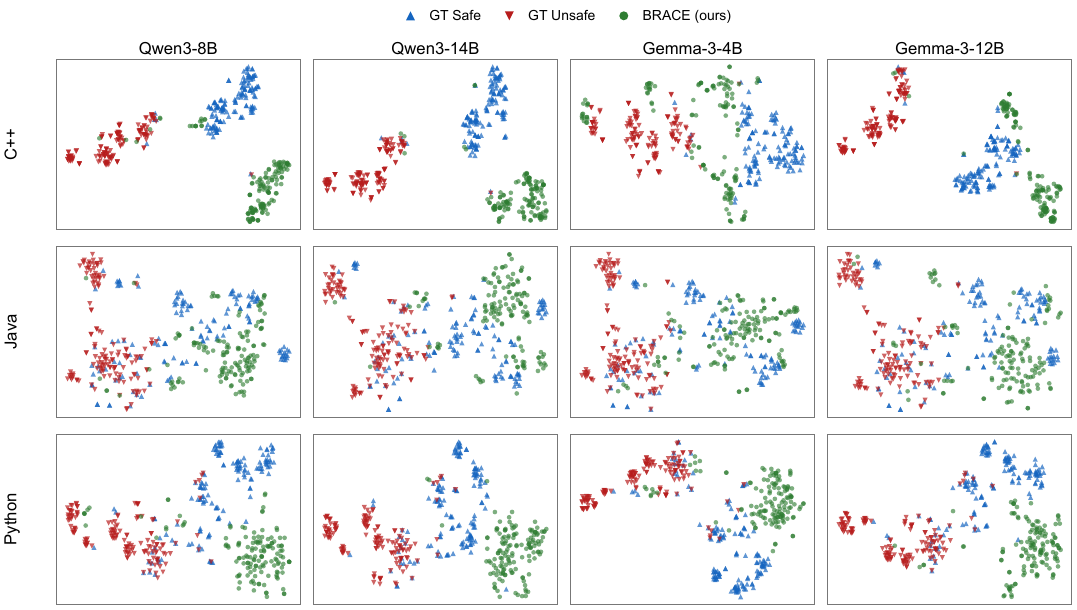}
    \caption{t-SNE projections of the judge's normalized last-layer representations for secure reference code, insecure reference code, and code generated by \ours{}. Each projection is fitted separately for one model--language setting using 142 samples from each group. The coordinates are meaningful only within a panel and should not be compared across panels.}
    \label{fig:composite_tsne}
\end{figure*}

The t-SNE projections in \autoref{fig:composite_tsne} provide a qualitative view of the same separation. Outputs from \ours often form their own subclusters rather than reproducing the secure reference cluster exactly. These subclusters usually occupy the secure side of the representation space and remain separated from the dense insecure regions. Their distinct position is consistent with implementation-level differences between generated and reference code, rather than evidence of a different security outcome.

\end{document}